\documentclass[twocolumn, 10pt]{IEEEtran}
\usepackage{graphicx}
\usepackage{amssymb}
\usepackage{amsmath}
\usepackage{mathtools}
\usepackage{dsfont}
\usepackage{cite}
\usepackage{stfloats}
\usepackage{subfigure}
\usepackage{psfrag}
\usepackage[mathscr]{euscript}
\usepackage{algorithm, algorithmic}
\usepackage{acronym}  
\usepackage{booktabs}
\usepackage{bbm}

\usepackage{multirow}

\usepackage{multicol}
\usepackage{adjustbox}

\usepackage{xcolor}

\DeclareUnicodeCharacter{FB01}{fi}

\usepackage{float}
\usepackage{balance}

\acrodef{CCDF}{complementary cumulative distribution function}
\acrodef{CF}{characteristic function}
\acrodef{PPP}{Poisson point processe}
\acrodef{RV}{random variable}
\acrodef{i.i.d.}{independent and identically distributed}
\acrodef{PDF}{probability distribution function}
\acrodef{CDF}{cumulative distribution function}
\acrodef{ch.f.}{characteristic function}
\acrodef{AWGN}{additive white Gaussian noise}
\acrodef{SNR}{signal-to-noise ratio}
\acrodef{LRT}{likelihood ratio test}
\acrodef{DRT}{distance ratio test}
\acrodef{GLRT}{generalized likelihood ratio test}
\acrodef{CRLB}{Cram\'{e}r-Rao lower bound}
\acrodef{CRB}{Cram\'{e}r-Rao bound}
\acrodef{ZZLB}{Ziv-Zakai lower bound}
\acrodef{ZZB}{Ziv-Zakai bound}
\acrodef{LOS}{line-of-sight}
\acrodef{ToF}{time-of-flight}
\acrodef{NLOS}{non-line-of-sight}
\acrodef{GDOP}{geometric dilution of precision}
\acrodef{GPS}{Global Positioning System}
\acrodef{FIM}{Fisher information matrix}
\acrodef{PEB}{position error bound}
\acrodef{SPEB}{squared position error bound}
\acrodef{TOA}{time-of-arrival}
\acrodef{TOF}{time-of-flight}
\acrodef{WSN}{wireless sensor network}
\acrodef{MAC}{medium access control}
\acrodef{RSS}{received signal strength}
\acrodef{WAF}{wall attenuation factor}
\acrodef{TDOA}{time difference-of-arrival}
\acrodef{RF}{radiofrequency}
\acrodef{RTT}{round-trip time}
\acrodef{AOA}{angle-of-arrival}
\acrodef{MF}{matched filter}
\acrodef{ED}{energy detector}
\acrodef{ML}{maximum likelihood}
\acrodef{MSE}{mean-square error}
\acrodef{RMSE}{root-mean-square error}
\acrodef{LEO}{localization error outage}
\acrodef{ppm}{part-per-million}
\acrodef{ACK}{acknowledge}
\acrodef{UWB}{Ultrawide bandwidth}
\acrodef{TNR}{threshold-to-noise ratio}
\acrodef{LS}{least squares}
\acrodef{IR-UWB}{impulse radio UWB}
\acrodef{FCC}{Federal Communications Commission}
\acrodef{TH}{time-hopping}
\acrodef{PPM}{pulse position modulation}
\acrodef{MUI}{multi-user interference}
\acrodef{PDP}{power delay profile}
\acrodef{BPZF}{band-pass zonal filter}
\acrodef{SIR}{signal-to-interference ratio}
\acrodef{SINR}{signal-to-interference-plus-noise ratio}
\acrodef{RFID}{radio frequency identification}
\acrodef{WPAN}{wireless personal area network}
\acrodef{WWB}{Weiss-Weinstein bound}
\acrodef{DP}{direct path}
\acrodef{MF}{matched filter}
\acrodef{MMSE}{minimum-mean-square-error}
\acrodef{SBS}{serial backward search}
\acrodef{SBSMC}{serial backward search for multiple clusters}
\acrodef{NBI}{narrowband interference}
\acrodef{WBI}{wideband interference}
\acrodef{INR}{interference-to-noise ratio}
\acrodef{CR}{channel response}
\acrodef{CIR}{channel impulse response}
\acrodef{CR}{channel  response}
\acrodef{RADAR}{radar}
\acrodef{MUR}{Multistatic radar}
\acrodef{JBSF}{jump back and search forward}
\acrodef{HDSA}{high-definition situation-aware}
\acrodef{RRC}{root raised cosine}
\acrodef{ST}{simple thresholding}
\acrodef{BTB}{Bellini-Tartara bound}
\acrodef{P-Max}{$P$-Max}  
\acrodef{MIMO}{multiple-input multiple-output}
\acrodef{MAP}{maximum a posteriori}
\acrodef{FG}{factor graph}
\acrodef{OP}{outage probability}
\acrodef{WED}{wall extra delay}
\acrodef{RMS}{root mean square}
\acrodef{SPAWN}{sum-product algorithm over a wireless network}
\acrodef{MDD}{minimum distance distribution}
\acrodef{MAP}{maximum a posteriori probability}
\acrodef{SAP}{small cell access point}
\acrodef{UE}{user equipment}
\acrodef{MBS}{macro cell base station}
\acrodef{UER}{\ac{UE} Relay}
\acrodef{D2D}{device-to-device}
\acrodef{MBS}{macro base station}
\acrodef{CSI}{channel state information}
\acrodef{OGR}{outage guard region}
\acrodef{FUR}{feasible UER region}
\acrodef{EHR}{energy harvesting region}
\acrodef{EH}{energy harvesting}
\acrodef{D2D-EHSN}{D2D communication provided \ac{EH} small cell network}
\acrodef{D2D-EHHN}{D2D communication provided \ac{EH} heterogeneous network}
\acrodef{3GPP}{3rd Generation Partnership Project}
\acrodef{BS}{base station}
\acrodef{DF}{decode and forward}
\acrodef{CCDF}{complementary cumulative distribution function}
\acrodef{ZF}{zero forcing}
\acrodef{RZF}{regularized zero forcing}
\acrodef{WLLN}{weak law of large number}
\acrodef{SLLN}{strong law of large numbers}
\acrodef{TDD}{Time-division duplex}
\acrodef{EE}{energy efficiency} 
\acrodef{HetNet}{heterogeneous network} 
\acrodef{SCP}{Single Cell Processing}
\acrodef{CBF}{Coordinated Beamforming}
\usepackage{color}
\usepackage{dsfont}
\usepackage{bbm}

\DeclareMathAlphabet{\mathsf}{OML}{cmbr}{m}{it}

\newcommand{\bd}{\begin{description}}
\newcommand{\ed}{\end{description}}
\newcommand{\be}{\begin{enumerate}}
\newcommand{\ee}{\end{enumerate}}
\newcommand{\bi}{\begin{itemize}}
\newcommand{\ei}{\end{itemize}}
\newcommand{\bl}{\begin{list}}
\newcommand{\el}{\end{list}}
\newcommand{\bt}{\begin{tabbing}}
\newcommand{\et}{\end{tabbing}}

\newcommand{\paperTitle}{The Role of Federated Learning in a Wireless World with Foundation Models}

\begin{document}



\title{\paperTitle}

\author{
    Zihan~Chen, \textit{Member, IEEE},
    Howard H. Yang, \textit{Member, IEEE},
    Y. C. Tay, \\
    Kai Fong Ernest Chong, \textit{Member, IEEE},
    and Tony Q. S. Quek, \textit{Fellow, IEEE}

\thanks{This paper is supported in part by the National Research Foundation, Singapore and Infocomm Media Development Authority under its Future Communications Research \& Development Programme, in part by the Ministry of Education, Singapore, under its Tier 2 Research Fund (MOE-T2EP20221-0016), and under its SUTD Kickstarter Initiative (SKI 2021\_03\_01), in part by the National Natural Science Foundation of China under Grant 62201504, in part by the Zhejiang Provincial Natural Science Foundation of China under Grant LGJ22F010001, in part by the open research fund of National Mobile Communications Research Laboratory, Southeast University (No. 2024D05), and in part by the Zhejiang – Singapore Innovation and AI Joint Research Lab. Y.C. Tay thanks SUTD for hosting his visit with ISTD. (\textit{Corresponding authors: T. Q. S. Quek, K. F. E. Chong})}

\thanks{Z. Chen, K. F. E. Chong, and T. Q. S. Quek are with the Singapore University of Technology and Design, Singapore 487372. (e-mail: \{zihan\_chen, ernest\_chong, tonyquek\}@sutd.edu.sg).}

\thanks{H. H. Yang is with the Zhejiang University/University of Illinois Urbana-Champaign Institute, Zhejiang University, Haining 314400, China, and also with the National Mobile Communications Research Laboratory, Southeast University, Nanjing 211111, China (email: haoyang@intl.zju.edu.cn).}

\thanks{Y. C. Tay is with the Department of Computer Science, National University of Singapore, Singapore 119077 (email: dcstayyc@nus.edu.sg)}

}
\maketitle
\acresetall
\thispagestyle{empty}
\begin{abstract} 
Foundation models (FMs) are general-purpose artificial intelligence (AI) models that have recently enabled multiple brand-new generative AI applications. The rapid advances in FMs serve as an important contextual backdrop for the vision of next-generation wireless networks, where federated learning (FL) is a key enabler of distributed network intelligence. Currently, the exploration of the interplay between FMs and FL is still in its nascent stage. Naturally, FMs are capable of boosting the performance of FL, and FL could also leverage decentralized data and computing resources to assist in the training of FMs. However, the exceptionally high requirements that FMs have for computing resources, storage, and communication overhead would pose critical challenges to FL-enabled wireless networks. In this article, we explore the extent to which FMs are suitable for FL over wireless networks, including a broad overview of research challenges and opportunities. In particular, we discuss multiple new paradigms for realizing future intelligent networks that integrate FMs and FL. We also consolidate several broad research directions associated with these paradigms.
\end{abstract}
\acresetall

\section{Introduction}\label{sec:intro} 
Foundation models (FMs), which include large language models (e.g. GPT series, LLaMA series, Gemini series, etc.) and large vision models (e.g. CLIP, SAM, etc.), are general-purpose artificial intelligence (AI) models that can be easily adapted to multiple downstream tasks~\cite{bommasani2021opportunities}. This adaptability is possible because of both size and scale:
FMs are sufficiently large models that display emergent abilities not found in smaller models~\cite{wei2022emergent}, and FMs are trained on massive (Internet-scale) data.
Recently, FMs have been the catalyst of multiple new AI applications.

Mirroring the success of AI, there has also been rapid progress in the development of intelligent wireless networks.
Indeed, network intelligence is envisioned to be a key component of the fifth-generation (5G)-and-Beyond wireless systems~\cite{chafii2023twelve}. Naturally, using FMs in wireless networks could also catalyze further research in network intelligence.

\begin{figure}[t!]
  \centering{\includegraphics[width=0.98\columnwidth]{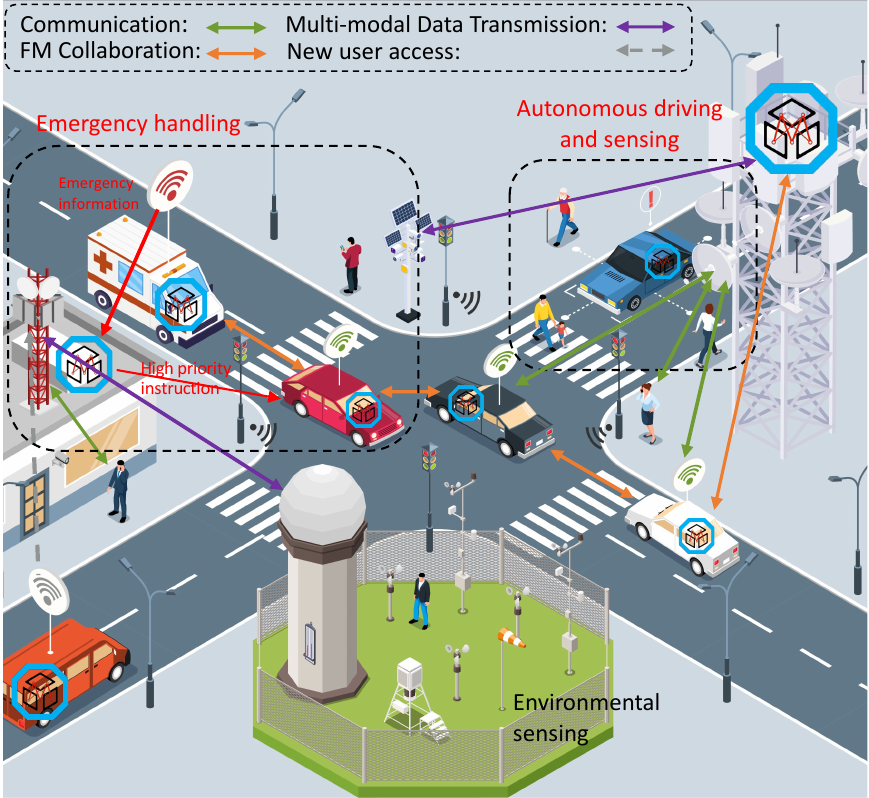}}
  \caption{An example of an intelligent transportation system over wireless networks, where the autonomous vehicles and edge server collaboratively perform low-latency decision-making assisted by FMs, such as route planning, traffic re-routing, and crowd management.}
  \label{fig:fm_trans}
\end{figure}

Imagine a world with intelligent infrastructure, where a transportation network is connected via an intelligent wireless network, with traffic lights and cameras, as well as autonomous vehicles connected to an FM-based AI system, in which decision-making such as re-routing traffic and managing crowds will be performed based on real-time monitoring of multi-modal data (e.g. weather, traffic, public events), together with tasks such as route planning and traffic sign understanding in FM-based autonomous driving; see Fig.~\ref{fig:fm_trans}. 
In this world, distributed network intelligence plays a central role, where multiple FMs across wireless networks could collaboratively work towards solving problems in real time. Hence, it is critical for such a highly autonomous but complex scenario to perform collaborations in FM training and inference, and multi-modal information fusion, so as to achieve low-latency and correct decision-making over wireless networks.

Federated learning (FL) is a concrete privacy-preserving paradigm for realizing distributed network intelligence, whereby the edge clients in the wireless network are involved in the deployment of intelligent services.
Such FL-enabled networks (which is also termed federated edge learning) offload the capabilities of intelligent prediction and decision-making to edge clients~\cite{MaMMooRam:17AISTATS}.
However, to effectively train a joint machine learning model across clients, challenges such as data heterogeneity and limited resources have to be addressed.
Numerous advanced FL approaches have been developed to enhance the performance of federated edge learning systems in terms of communication efficiency, robustness, and personalization~\cite{wang2021field}.

In a world with FMs, how do we reconcile FMs and FL over wireless networks to advance distributed network intelligence? Such an interplay between FMs and FL is naturally a confluence of opportunities and challenges.
Intuitively, FL can be used to train FMs in a distributed manner, where computing resources and private data can be used in the training stage, thereby complementing the usual centralized training scheme.
Dually, the excellent adaptability of pre-trained FMs could benefit the different (pre-processing/training/inference/evaluation) stages of federated edge learning systems and strengthen the intelligent decision-making ability of future wireless networks. 
Such opportunities are accompanied by technical challenges.
Compared to conventional machine learning models,
the training and inference processes of FMs are cost-intensive across various aspects: memory, storage, computing, and communication overhead. The involvement of human feedback to further fine-tune FMs is also a crucial step for some FMs~\cite{ouyang2022training}. 
When considering the deployment of FMs over a wireless network, either at the edge or in the cloud, for the purpose of either training or inference, the inherent nature of FMs poses critical challenges to storage, power consumption, and communication traffic.

In this article, 
we shall present a balanced discussion on both \textbf{opportunities} and \textbf{challenges}, 
by answering the following question:\\
\textbf{\textsf{To what extent are FMs suitable for FL-enabled wireless networks?}}

\noindent In particular, we focus on the following sub-questions:
\textit{How to design a sustainable paradigm for deploying systems that integrate FM and FL?}  
\textit{What are the key challenges and constraints?}
To explore the possible new paradigms as well as the potential challenges for the interplay between FMs and FL, we provide a broad overview of the deployment of FMs over a federated edge learning system in Sec.~\ref{Sec:challenges}. 
In Sections ~\ref{Sec:FM_in_FL} and \ref{Sec:FL_for_FM}, we address in detail how FMs and FL could benefit each other. We also highlight some future directions.

\section{Paradigms, challenges, and opportunities for the integrated FM and FL}\label{Sec:challenges}
We begin this section with a brief introduction to FMs and FL. Possible paradigms of integrated FL and FMs will be presented subsequently.

\begin{figure}[t!]
  \centering{\includegraphics[width=0.91\columnwidth]{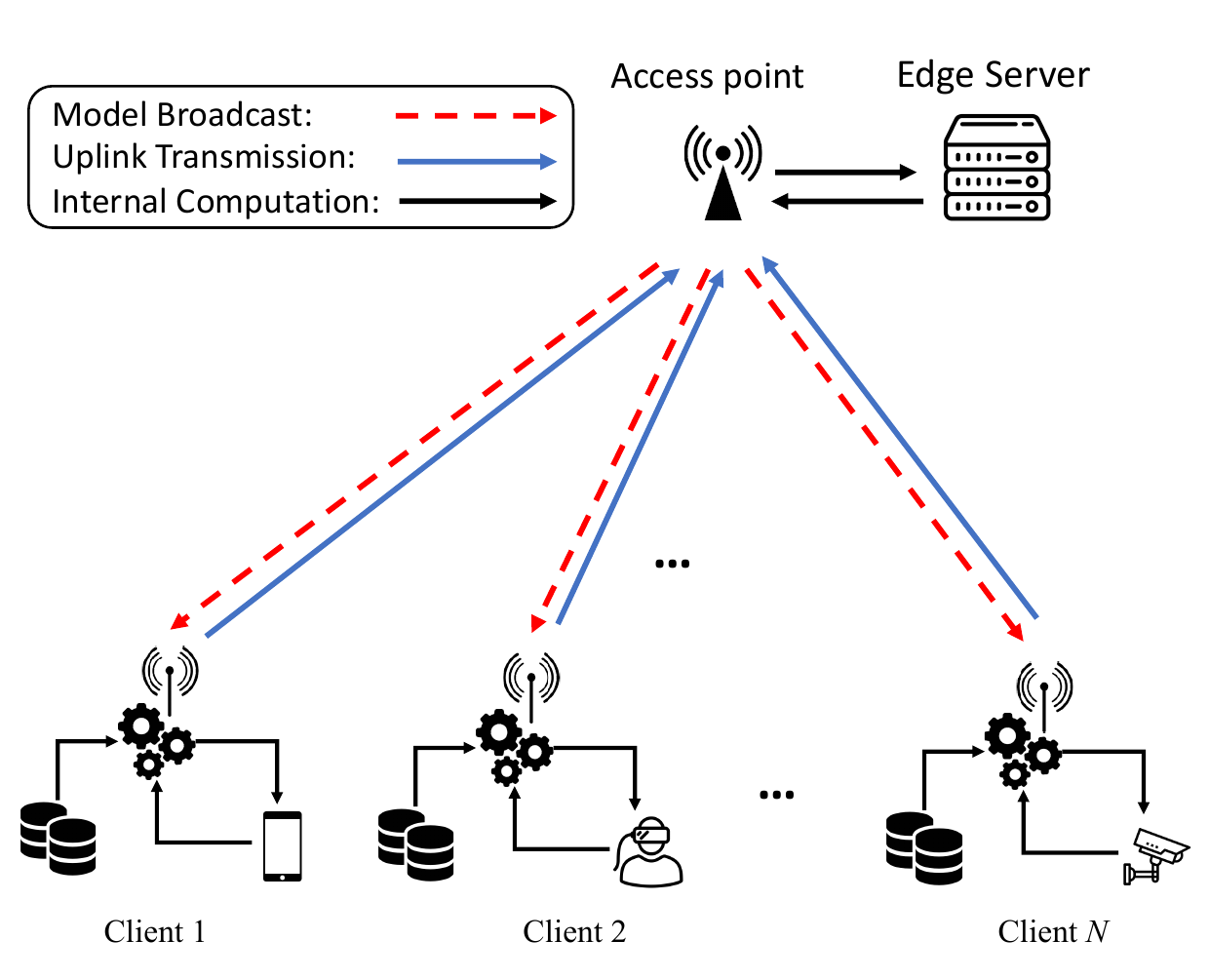}}
  \caption{A brief overview of vanilla machine learning paradigm in FL-enabled wireless networks.}
  \label{fig:fl}
\end{figure}

\subsection{Preliminaries of FMs and FL} 
\noindent\textbf{Foundation models.} In this article, FMs refer to a class of models trained over huge amounts of data and consist of parameters in the billions range, demonstrating emergent capabilities across applications and tasks like linguistic, visual, robotics, and reasoning (See Tab.~\ref{tab:summary} for a representative summary of FMs~\cite{bommasani2023holistic}).
A typical training pipeline of FMs consists of self-supervised training, supervised fine-tuning, and reinforcement learning with human feedback (RLHF). 
Overall,  the level of access (e.g. via paid API access or downloadable as open-source models) is the primary consideration for how FMs could be integrated into wireless networks.
For instance, usual training and customized fine-tuning could be carried out with open-source FMs, which may not be possible for proprietary FMs.

\noindent\textbf{Federated learning.} FL is a distributed learning paradigm in which multiple clients collaboratively train a machine learning model, coordinated by a server without exposing local data. (See Fig.~\ref{fig:fl} for a brief overview.) The typical federated training process is organized in terms of communication rounds, where the model parameters are exchanged between the clients and the server.
Broadly speaking, FL could be divided into two general types: \textit{cross-device FL} (i.e. training over a large number of clients, typically with limited training data) and \textit{cross-silo FL} (i.e. collaborative training with a limited number of clients, e.g. hospitals and companies, typically with a large amount of training data). In both types, the challenges of limited communication and system/data heterogeneity would have to be addressed~\cite{wang2021field}.

\begin{table}[t!]
\caption{A comparison of open/closed-source foundation models, where the energy and hardware costs are with respect to model pre-training.}
\begin{tabular}{l|clcc} \toprule
Models & Open  & \# of paras.  & Energy cost  & Hardware cost   \\ \midrule
GPT-3 &  &175B  &1287 MWh  & -   \\  
PaLM-2 &  &340B  & -   & - \\ 
LLaMA-2 & $\checkmark$  &7B-70B  &2638 MWh  & 3.3M GPU hours \\ 
BLOOM &$\checkmark$  &176B  &  433 MWh  & $\approx$1M GPU hours \\ 
Falcon 180B & $\checkmark$   &180B  & -  & $\approx$7M GPU hours \\ \bottomrule
\end{tabular}
\label{tab:summary}
\end{table}

\subsection{Challenges of deploying FMs over FL-enabled wireless networks}
The FL-enabled wireless network consists of multiple local clients and an edge server. 
Before integrating FMs into an FL-enabled wireless network, we have to take into consideration the conflicts arising from the constraints imposed by different FL scenarios versus the cost-intensive requirements of using FMs.
We summarize the key challenges as follows.

\begin{itemize}
    \item \textit{High power consumption.} Due to the sheer Internet-scale data required for training and the tremendous number of model parameters, the training process of FMs will have substantial requirements on both computation hardware and energy consumption (See Tab.~\ref{tab:summary}~\cite{bommasani2023holistic}).  
    This makes the role of energy efficiency even more critical when deploying energy-hungry FMs in a reasonably sustainable manner.
    Moreover, to integrate FMs into wireless networks, we typically require specialized hardware for AI computing acceleration, such as GPUs and TPUs.
    \item \textit{Large storage and memory requirements.} To satisfy the requirements of training FMs, both storage and memory would have to be drastically increased to handle streaming collected/generated data and the updating of model parameters during training.
     Typical network architectures may not necessarily meet such storage and memory requirements. 
    \item \textit{Huge communication overhead.} 
    Due to the sheer model sizes of FMs, communication overhead would be immense even for transmitting pre-trained/fine-tuned FM weights from the FM vendor to the edge of the network for downstream tasks. For FL-based training, since training FMs from scratch could take up to six months of continuous training over thousands of GPUs, this communication overhead would be magnified in both downlink and uplink over a sustained period of time.
    \item \textit{Additional latency.} The 5G-and-Beyond network has strict latency requirements. 
    It is not clear how the integration of FMs into FL-enabled networks would meet such strict latency requirements.
    \item \textit{Hallucination of FMs.} 
    Hallucination is a crucial and inevitable challenge when using FMs~\cite{bommasani2023holistic}. Here, hallucination refers to the generation of incorrect, fabricated, or nonsensical information by FMs. This could lead to disastrous outcomes, especially if FMs are relied upon for critical automated decision-making over wireless networks, such as tasks in autonomous driving. 
    Once FMs are deployed, it becomes a challenge how the negative outcomes of hallucinations could be detected and alleviated for \textit{correct} decision-making over wireless networks.
\end{itemize}

\begin{figure*}[t!]
\centering{\includegraphics[width=0.92\linewidth]{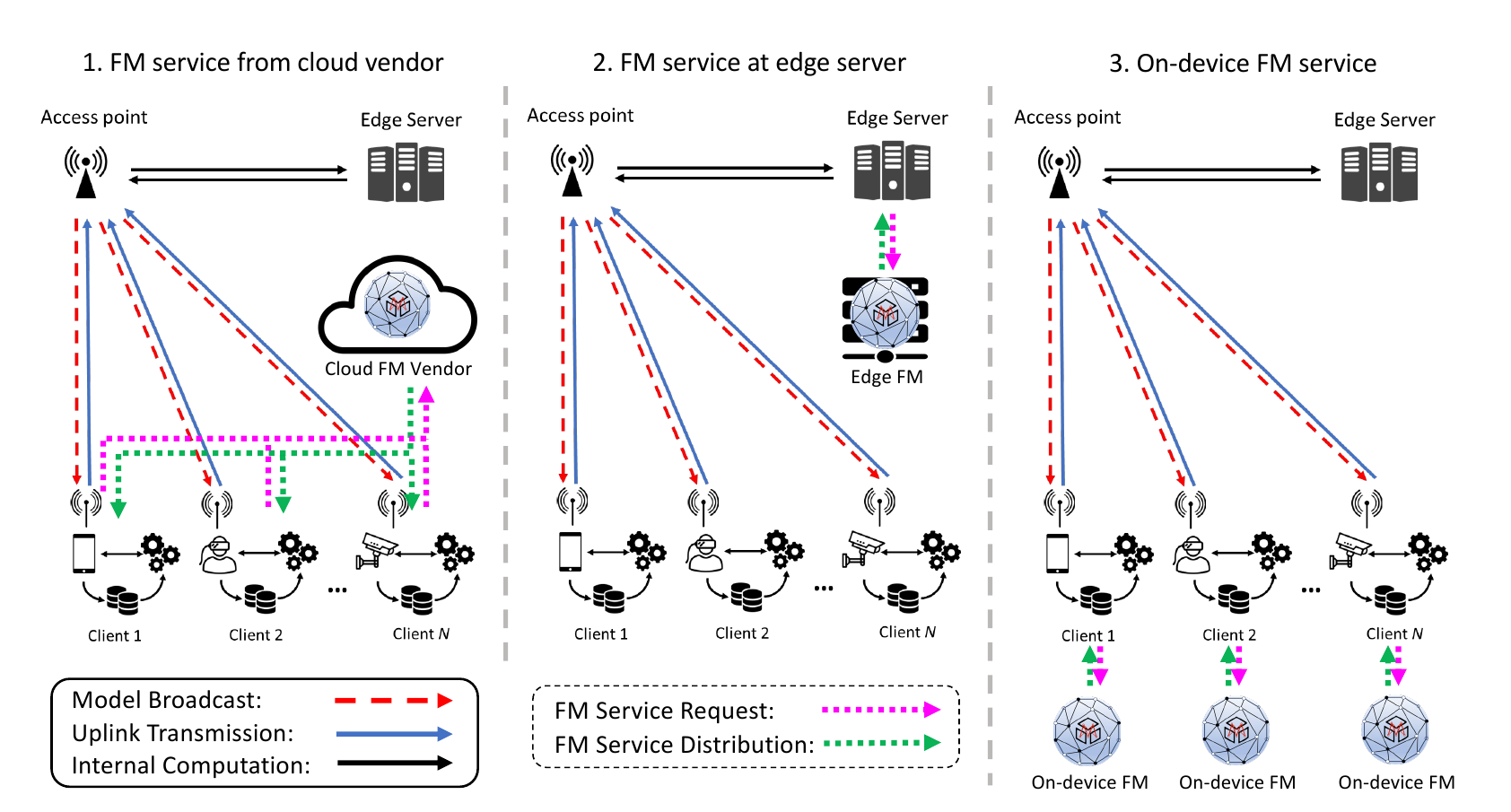}}
  \caption{An overview of different types of architecture of integrated FMs in the federated edge learning system. FMs are deployed at the cloud server (1), edge server (2), and local clients (3), respectively.}
  \label{fig:fm4fl}
\end{figure*}

\subsection{Possible network architectures for integrating FM and FL}
Despite the exciting possibilities brought by FMs, the aforementioned cost-intensive challenges that FMs pose would naturally constrain the architectural design for a practical, intelligent wireless network that integrates FMs and FL. For example, the stringent requirements for storage, memory, and computing resources would prevent FMs from being deployed at the edge of networks in cross-device FL settings. 
Taking into consideration such constraints and the goal of achieving sustainable integration of FL and FMs, we summarize the possible architectures as follows, with respect to how FMs and FL-enabled networks could benefit from each other.

\noindent\textbf{FMs in FL-enabled wireless networks.}
In real-world FL-enabled wireless networks, 
the inherent system heterogeneity is reflected by the diverse computation and communication capabilities across different clients, as well as the server. Hence, we cannot hope to have a ``one-size-fits-all'' scheme for the position and function of FMs. 
Specifically, for a system with sufficient computing and hardware resources at the edge clients, each client could afford a generic/personalized FM. 
In contrast, for a resource-constrained system, we could have an FM deployed at the edge server, or an FM provided by a cloud vendor via paid API access.

\noindent\textbf{FL for training FMs over networks.}
In resource-constrained wireless networks, the usual FL model updates and aggregations may not be feasible due to the immense computing and communication requirements for training FMs.  
This makes a hybrid global-local (cloud-edge) model training scheme more preferable.
In particular, we could offload cost-intensive computations (e.g. FM pre-training) to the server, and reserve lower-cost computations (e.g. fine-tuning/personalization) for the clients, to alleviate both communication overhead and the high demand for computational resources. 
Furthermore, parameter-efficient tuning strategies could also be adopted to achieve collaborative training across clients with minimum costs.

Detailed descriptions of both scenarios will be provided in Sections~\ref{Sec:FM_in_FL} and \ref{Sec:FL_for_FM}, respectively.

\section{FMs in FL-enabled wireless networks}\label{Sec:FM_in_FL}
Consider an FL-enabled network with $N$ clients. 
The limited communication resources and the inherent data/system heterogeneity of the real-world networks would hinder the performance and scalability when deploying FL. 
With the inclusion of FMs as a critical component of our intelligent network infrastructure, we could leverage FMs to enhance FL training performance and provide new application scenarios that conventional AI models cannot provide. 

However, as indicated previously, there is no ``one-size-fits-all'' scheme. The way that FMs are integrated into an FL system should be aligned with the system's properties,
where conceptually, FMs play the role of a customized service provider.
In other words, we can think of the usage of FMs abstractly in terms of 
``\textit{Foundation Model as a Service}'' (FMaaS).
Specifically, in existing FL-enabled networks, we could use FMs to provide different types of services in different stages, such as data pre-processing, training, and the calibration of the jointly trained model, described as follows.

\subsection{Foundation models in pre-processing stages}
The imbalanced data widely existing in real-world networks, and the induced data heterogeneity across clients is regarded as a major challenge that constrains the training performance of FL systems. 
As one of the most widely known properties of some FMs,  the generative abilities can be applied to enhance the model training~\cite{du2023enabling,schick2021generating}.
Due to the high-cost property, FM could be deployed either at the edge server or at the cloud data center to facilitate federated training of small models over wireless networks, as described in the first two paradigms in Fig.~\ref{fig:fm4fl}.
Possible integrated systems are summarized as follows.

\noindent\textit{1) Data augmentation at the edge:} The imbalanced data in heterogeneous networks may result in a limited number of data samples in a few classes (e.g., the minority classes in classification and recognition tasks), leading to poor representation capabilities of small models in these classes. 
In this context, clients with local imbalanced data at the network edge could make a request to FM for supplementary data generation in minority classes, to achieve balanced local data statistics for balanced local representation learning.

\noindent\textit{2) Synthetic data at the server:}
In the case of service unavailability and outage from the FM vendor (due to poor connection or 
traffic congestion), a synthetic (balanced) dataset could be constructed at the edge server based on the generative services of FMs. Given the synthetic data, multiple scenarios could be explored. Firstly, at the end of each round, the data could be used to evaluate the aggregated model or the local models separately for post-processing (e.g. model re-weighting) to improve the model generalization performance and robustness, as the statistics of the model weights may be diverse in the presence of data heterogeneity or adversarial attacks. 
Secondly, the new dataset could play a role in model distillation at the server, in which a pre-trained FM could be regarded as the teacher model. The post-training at the server could utilize these data for global model calibration and feature alignment.

The role of FMs as data augmenters would get the public or local synthetic data involved in global distillation or local training, enabling small models to learn more balanced representation from combined datasets rather than local private data. Therefore, it would significantly improve the performance of privacy protection and the robustness to adversarial attacks.
In summary, the FM services of auxiliary data augmentation/generation could not only mitigate the negative effects brought by imbalanced data in FL but also improve the privacy protection performance of data.

\subsection{Foundation model in the training}
In addition to the generative ability, the fast adaptation ability of FMs could be leveraged as a tool that could be involved in the training process. In this subsection, we will discuss a few applications that leverage FMs to assist in training robust (and relatively small) models over FL-enabled wireless networks~\cite{shridhar2023distilling}.
 
In conventional deep learning, a well-trained model could act as a teacher model to transfer the knowledge for small model training (i.e. the student model).
Since the pre-trained FMs acquired huge amounts of knowledge from the massive training data, it is natural to design an integrated system to retrieve and transfer the acquired knowledge of FMs to boost the small model training over FL-enabled networks.
However, the deployment strategies may differ in systems with diverse hardware capabilities.
\begin{itemize}
    \item In cross-device FL-enabled networks, it is difficult to conduct on-device training for FMs.\footnote{On-device FM deployment may be feasibly accomplished through the adoption of efficient FM deployment techniques~\cite{lin2023awq}.} The server-side model deployment becomes an alternative practical option, in which the clients could request corresponding FM services during training via cloud-edge collaboration over wireless networks. The induced additional latency and communication overhead over wireless networks could be alleviated by adopting quantization~\cite{lin2023awq}, specific communication mechanism~\cite {zhao2023lingualinked}, and other latency management approaches~\cite{chafii2023twelve}. For example, FMs could take the role of teacher models in knowledge distillation to enhance the training of small models~\cite{shridhar2023distilling}, which is a typical application of FM-based cloud-edge collaboration to distill the capabilities of FMs into smaller models.
    \item In cross-silo FL or edge devices with powerful hardware (e.g. autonomous vehicles), FMs could be deployed locally, in which the pre-trained FMs could be involved in training to boost the performance of the local model. Specifically, techniques such as transfer learning and knowledge distillation could help the local model achieve better generalization under the guidance of FM with superior knowledge extraction and representation learning capabilities; see the third paradigm of Fig.~\ref{fig:fl4fm}.

\end{itemize}

\subsection{Foundation models for model evaluation:}
As discussed previously, in cross-device FL-enabled networks, deploying FMs for local training may be unrealistic. Alternatively, FMs at the edge server could be opted with more functionality with access to the updated local models or aggregated model, rather than only being involved in the training.
Since the pre-trained FMs exhibit excellent performance on multiple downstream tasks, the performance of FMs could be regarded as a benchmark for evaluating the smaller models. The current performance evaluation and validation are only based on limited validation and test datasets. Although the trained model achieves good generalization performance over the test set, there still exist potential over-fitting risks. 
Hence, the output of FM could be used as a criterion of performance evaluation for a smaller model by comparing it with the corresponding output.
Once the pre-trained FMs are available in FL-enabled networks, the edge server could request such services for complementary model evaluation.

\subsection{Potential issues}
Leveraging FMs brings promising opportunities to the FL-enabled network design and the training frameworks.
To integrate FM into the FL-enabled networks, a few properties of the wireless networks and FMs would still constrain the deployment and the scalability of the integrated systems.
To meet the requirements of future intelligent networks, the following issues need to be addressed.

\noindent\textit{1) Continuous and stable wireless connection services:} 
The time-varying property of wireless channels may lead to occasional unstable connections. It is possible that a service outage occurs among the clients, edge servers, and FM vendors. For example, the mobility property in autonomous driving brings challenges to service continuity. 
On the other hand, low latency is a key consideration of wireless communication systems. The latency brought by the FM service request may degrade the performance especially when the task outputs are involved in the latency-aware decision-making process during local or global model training/evaluation.
Strategies including model quantization, tensor parallelism, and operator fusion could be explored for latency performance enhancements~\cite{lin2023awq}.

\noindent\textit{2) FM Service alignment and security:} 
Despite the superior capabilities across a vast expanse of scenarios, the FM performance may fluctuate in real-world applications, with respect to fairness, bias, toxicity, and other related metrics~\cite{bommasani2023holistic}.  The integrated system design must take these possible uncertainties into consideration, to improve the robustness and sustainability of the integrated systems.
Furthermore, it is also known that the total time for FM inference varies significantly among different tasks. Similar to the first point, diverse latency would be newly introduced during this process.  
Thus, a joint design of the computation and communication for low latency should be explored to ensure the overall performance~\cite{zhao2023lingualinked}.

\section{FL for training FMs over networks}\label{Sec:FL_for_FM}
\begin{figure*}[t!]
\centering{\includegraphics[width=0.90\linewidth]{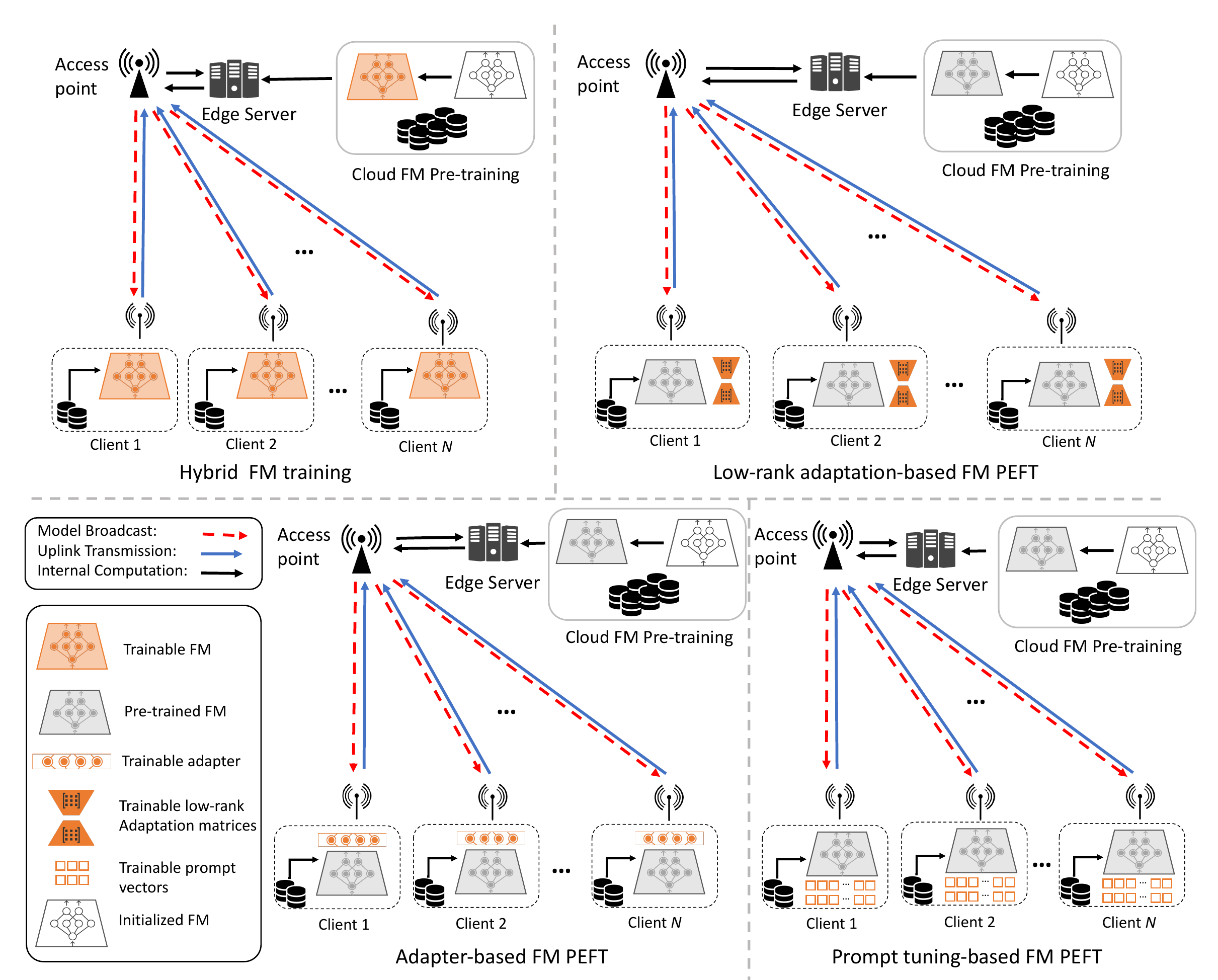}}
  \caption{An overview of different types of training paradigms of FMs in the federated edge learning system with cloud FM pre-training. In PEFT-based paradigms, the parameters of pre-trained FMs, depicted as gray blocks,  will remain intact, and only a small portion of parameters (e.g., adapter weights and prompt vectors) will be updated and transmitted.}
  \label{fig:fl4fm}
\end{figure*}

For the training of FMs of increasingly larger scales, after all publicly available data has been exhausted as training data, the next frontier would inevitably be personal data, which is naturally distributed across wireless networks.
Due to the inherent privacy issues arising from the use of personal data, it is frequently not feasible to aggregate personal data from multiple sources into a dataset for centralized training.  
Hence, FL is well-positioned in this new frontier, for the privacy-preserving training of FMs on personal data.
Moreover, the distributed nature of FL allows for the use of computing resources across networks that could otherwise be idle, although we still have to address the issue that such computing resources may be limited. 
In this section, we discuss some possible scenarios for training FMs over FL-enabled networks.\footnote{Note that we only consider cross-silo FL (with adequate infrastructure for sustained energy consumption and sufficient local data for FM training). Cross-device FL naturally cannot be compatible with the local deployment and training of FMs over wireless networks. 
The scheme of usual federated training from scratch will also not be included due to the unexplored critical challenges of integrating unsupervised learning and distributed RLHF into FL.} 
In particular, we will discuss pre-trained FM-based hybrid training and parameter-efficient training; see Fig.~\ref{fig:fl4fm}.

\subsection{Hybrid training for FM over networks}
As discussed previously, the usual FL training for FM brings uncertainty and challenges.  
In centralized scenarios, the pre-trained FM has become the bedrock for mainstream applications. Motivated by this, we shall consider a ``pre-train globally, fine-tune locally'' scheme. More specifically, the overall training process for FMs shall be split into two parts, in which the cloud or edge server will be in charge of the pre-training in the first stage to obtain a pre-trained FM via utilizing the centralized public data and then the distributed clients will be involved in the subsequent training process for further local full model fine-tuning or personalization based on the local private data.

The hybrid training scheme could be further divided into different types according to different objectives in the second stage. 
For generic training purposes (i.e., to train a shared FM for all clients), a short FL training stage will be coordinated across the clients after receiving the pre-trained FMs, to enhance the performance of shared FMs, which would be obtained via the usual model aggregation in FL. 
For personalized training purposes (i.e., each client maintains an FM), each local client will train its own personalized FM via personalized FL training frameworks. The personalization degree could be adjusted according to local preferences.

\subsection{Parameter-efficient fine-tuning}
No matter whether in training from scratch or in local fine-tuning based on the pre-trained FM, all the parameters of FMs will be updated. For large-scale training, such schemes may be infeasible due to the excessive training. The full model parameter update process, including forward and backward propagation, would incur additional memory and storage costs for gradients and other intermediate parameters. 
Enormous efforts have been devoted to addressing such constraints with low-cost solutions~\cite{hu2021lora,jiang2023low}, among which parameter-efficient fine-tuning (PEFT) is proposed to achieve fast adaptations with much less trainable parameters. Compared to the other schemes, performing PEFT at the local clients could be regarded as a fast adaptation method for deploying local FMs in the FL-enabled networks. In the following, we discuss how PEFT could benefit the training of FMs over FL-enabled networks.

\begin{itemize}
    \item \textit{1) Adapter-based PEFT in FL:} The adapter-based PEFT introduces an additional adapter built upon the pre-trained model, which consists of a few layers with a small amount of trainable parameters. The adapter is usually inserted between existing layers or follows after the output layer of FMs. For the combined structure, only the adapter is updated in PEFT. The number of trainable parameters and the cost of computing and storage could be vastly reduced. Also, in FL, such integration means that only the parameters of the adapter will be transmitted for model aggregation over wireless networks. The corresponding communication cost will also be much smaller. Nonetheless, additional inference latency will be induced due to the extended data path brought by the adapter layer.
    \item \textit{2) Low-rank adaptation-based PEFT in FL:} Unlike the inserted adapter, the low-rank adaptation methods (e.g. LoRA) merge trainable low-rank matrices in parallel to the frozen pre-trained weight matrices of FMs. Benefited by the parallel merging, the low-rank adaptation methods do not introduce additional inference when compared to the pre-trained FM~\cite{hu2021lora}. Aggregation would also be conducted only among the low-rank matrices with low communication costs over wireless networks. To provide personalized services, personalized adaptation matrices could also be considered.
    \item \textit{3) Prompt tuning-based PEFT in FL:} The adaptation methods demonstrate effectiveness on downstream tasks but trainable parameters still require frequent parameter exchange with potential privacy concerns. Prompt tuning (or prompt engineering) provides another perspective to improve the FM performance\footnote{Prompt consists of tunable tokens per downstream task to be prepended to the input text, which can be regarded as learnable parameters.}. The interplay between prompt tuning and FL enables federated prompt tuning for communication-efficient solutions~\cite{guo2023promptfl}, where only the prompt vectors are learned and shared. As no parameters of the FMs are trained, the prompt-tuning method provides promising solutions to realize communication-efficient and sustainable networks. 
\end{itemize}

\begin{table*}[t!]
\caption{Performance comparison of different PEFT methods over WikiText-2 dataset with respect to fine-tuning time, latency, and perplexity, over two setups: with and without quantization.  All experiments are conducted over an FL system with 20 clients with random partitioned data. Fine-tuning time refers to the averaged local fine-tuning time per client in a single communication round. The latency given in the table indicates the inference latency per token. Perplexity is a well-studied metric for language tasks, where a lower perplexity means better model performance.}
\centering
\begin{adjustbox}{width=1.7\columnwidth,center}
\begin{tabular}{l|cccccc} \toprule
\multirow{2}{*}{Setup}   & \multicolumn{3}{c|}{w/o quantization} & \multicolumn{3}{c}{w/ quantization}\\\cline{2-7}
   & Fine-tuning time &  Latency & Perplexity  & Fine-tuning time &  Latency & Perplexity  \\ \midrule
LLaMA-7B + Adapter  & 3.4 min  & 25.3 ms  & 6.17  & 3.1 min  & 22.3 ms  & 6.38 \\  
LLaMA-7B + LoRA & 3.5 min  & 23.9 ms  & 6.23 & 3.1 min  & 21.6 ms  & 6.29 \\  \bottomrule
\end{tabular}
\end{adjustbox}
\label{tab:latency_compare}
\end{table*}

\textbf{Case study.}
A comparison of LLaMA-7B with adapter versus LLaMA-7B with LoRA over an FL system is illustrated in Tab.~\ref{tab:latency_compare}, where we compare the training performance of these two PEFT methods with/without AWQ~\cite{lin2023awq} 4-bit quantization. Note that with this quantization, the overall weight memory cost is decreased from 13.6 GB to 3.7 GB. As our experimental results demonstrate, AWQ quantization is able to achieve FM compression over FL systems, reducing memory cost and latency while maintaining similar perplexity.

In summary, training FMs over wireless networks could leverage the distributed data and power in a privacy-preserving manner but also present cost-intensive drawbacks with respect to the computation, storage, and communication resources. As our case study suggests, PEFT in conjunction with strategies such as AWQ quantization, provides a possible solution for on-device deployment and efficient training of FMs over FL-enabled networks. 
We believe further work should be done to address the potential challenges in deployed FMs.

\section{Future Trends and Open Issues}
We have discussed the challenges of integrating FMs with FL-enabled networks, and also highlighted how FMs and FL could benefit each other over wireless networks. To exploit the potential gains and address the challenges for the interplay in a robust and sustainable manner, 
we propose the following broad areas that warrant further investigation. 
\begin{itemize}
    \item \textit{Incentive design for FM service request and client participation.} Training FMs is costly. In FM-integrated FL-enabled networks, the request for FM services (e.g. synthetic data generation and pre-trained FM downloading) may not be granted by the FM vendor without a proper incentive mechanism. 
    How we incentivize the FM vendor to provide FMaaS remains an open question. 

    \item \textit{Joint optimization of FM QoS and computing resources.} 
    As one of the key characteristics of FM training, the huge computational demands for training, inference, storage, and communication pose a critical challenge in resource-constrained wireless networks. It is important to efficiently utilize the limited resources across the networks while ensuring the QoS of FM services. A joint optimization scheme to balance the trade-off between the QoS of FM services and the resources in FL-enabled networks could be further explored.
    
    \item \textit{Privacy and robustness issues for FM services.} In AI-related tasks and services, 
    the preservation of both data privacy and model privacy are increasingly important factors for building trustworthy AI systems.
    Numerous privacy-preserving mechanisms and robustness to different adversarial attacks have been investigated in FL-enabled networks. However, it is not clear how such methods would perform in a wireless world of FMs. 

    \item \textit{Task scheduling for low-latency services.} Latency is a key consideration in designing intelligent networks.
    In FM-integrated FL-enabled networks, the non-negligible latency induced by the integration of FMs is heavily task-dependent.
    We foresee that task-adaptive scheduling protocols would be increasingly important as FMs become more prevalent in future wireless networks.

    \item \textit{Communication protocol design for transmission of FMs.} On the road towards ubiquitous intelligent networks, the transmission of FM weights, as illustrated in Fig.~\ref{fig:fm4fl} and Fig.~\ref{fig:fl4fm}, would be a non-negligible component of network traffic. Currently, there is no specific coding and communication protocol for the transmission of FMs. It would be necessary to develop efficient protocols and coding techniques for FM transmission while considering the inherent structural properties of FM architectures, data flow between FM segments, as well as model integrity.

\end{itemize}

\section{Concluding Remarks}
The rapid advances in FMs serve as an important contextual backdrop for the vision of FL-enabled intelligent wireless networks, while the exploration of the interplay between FMs and FL is still in the nascent stage.
In this article, to explore the extent to which FMs are suitable in FL-enabled wireless networks, we presented a balanced discussion on both opportunities and challenges. A broad overview of the possible paradigms of the integrated FMs and FL was exploited.
We finally provided potential future trends for achieving robust and sustainable integration of FMs and FL over wireless networks.

\bibliographystyle{IEEEtran}
\bibliography{bib/StringDefinitions,bib/IEEEabrv,bib/fl_fm}

\begin{IEEEbiographynophoto}
    {Zihan Chen} received the B.Eng. degree in Communication Engineering from the Yingcai Honors College at the University of Electronic Science and Technology of China (UESTC) in 2018. He received his Ph.D. degree from the Singapore University of Technology and Design (SUTD)-National University of Singapore (NUS) Joint Ph.D. Program in 2022. Currently, he is a Postdoctoral Research Fellow at SUTD.  His research mainly focuses on network intelligence, machine learning and edge computing.
\end{IEEEbiographynophoto}

\begin{IEEEbiographynophoto}
    {Howard H. Yang} received the Ph.D. degree from Singapore University of Technology and Design (SUTD), Singapore, in 2017. He was a Postdoctoral Research Fellow at SUTD from 2017 to 2020, he also had research attachments with Princeton University and The University of Texas at Austin. Currently, he is an assistant professor at the ZJU-UIUC Institute, Zhejiang University, Haining, China. Dr. Yang currently serves as an editor for the {\scshape IEEE Transactions on Wireless Communications}. He received the IEEE ComSoc Asia-Paciﬁc Outstanding Young Researcher Award in 2023, the IEEE Signal Processing Society Best Paper Award in 2022, and the IEEE WCSP 10-Year Anniversary Excellent Paper Award in 2019.
\end{IEEEbiographynophoto}

\begin{IEEEbiographynophoto}
    {Y. C. Tay} received his BSc degree from the University of Singapore and his Ph.D. degree from Harvard University. He is a professor in the Department of Computer Science. His main research interests are performance modelling (caching systems, wireless protocols, Internet traffic, database transactions) and database systems (synthetic generation of data, social networks and concurrency control).
\end{IEEEbiographynophoto}

\begin{IEEEbiographynophoto}
    {Kai Fong Ernest Chong} graduated summa cum laude from Cornell University with a B.A. in Mathematics in 2009, and he also received his Ph.D. in Mathematics from Cornell University in 2015. He was the recipient of the 2003 Singapore National Science Talent Search grand prize, and the 2015 Eleanor Norton York Award. Currently, he is an assistant professor at Singapore University of Technology and Design (SUTD). He serves as the co-chair of his department's undergraduate studies committee, and he serves as SUTD's thrust lead for the "Fundamentals and Theory of Artificial Intelligence (AI) Systems", which is a thrust within the AI/Data Science sector of SUTD's growth plan in research and education. His research spans both Mathematics and AI. Within AI, he is interested in algebraic methods in AI, theoretical and computational aspects of deep learning, machine reasoning, and robustness for AI.
\end{IEEEbiographynophoto}

\begin{IEEEbiographynophoto}
    {Tony Q.S. Quek}(S'98-M'08-SM'12-F'18) received the B.E.\ and M.E.\ degrees in electrical and electronics engineering from the Tokyo Institute of Technology in 1998 and 2000, respectively, and the Ph.D.\ degree in electrical engineering and computer science from the Massachusetts Institute of Technology in 2008. Currently, he is the Cheng Tsang Man Chair Professor, ST Engineering Distinguished Professor, and Head of ISTD Pillar with Singapore University of Technology and Design as well as the Director of Future Communications R\&D Programme. He was honored with the 2008 Philip Yeo Prize for Outstanding Achievement in Research, the 2012 IEEE William R. Bennett Prize, the 2017 CTTC Early Achievement Award, the 2017 IEEE ComSoc AP Outstanding Paper Award, the 2020 IEEE Communications Society Young Author Best Paper Award, the 2020 IEEE Stephen O. Rice Prize, the 2020 Nokia Visiting Professor, and the 2022 IEEE Signal Processing Society Best Paper Award. He is a Fellow of IEEE and a Fellow of the Academy of Engineering Singapore.
\end{IEEEbiographynophoto}

\end{document}